\documentclass[aps,prd,10pt,preprint,superscriptaddress,onecolumn,nofootinbib]{revtex4-1}
\usepackage{graphicx, epsfig} 
\usepackage{amsmath,amssymb,amsfonts,dsfont,mathrsfs,amsthm,mathtools}
\usepackage{bm} 
\usepackage{color}
\usepackage[usenames]{xcolor}
\usepackage{hyperref}
\usepackage{verbatim}
\usepackage{siunitx}
\hypersetup{linktocpage,colorlinks=true,urlcolor=blue,linkcolor=magenta,citecolor=blue}
\usepackage[normalem]{ulem}

\newcommand{\stkout}[1]{\ifmmode\text{\sout{\ensuremath{#1}}}\else\sout{#1}\fi}

\newcommand{\diff}[1]{\text{d}#1}

\newcommand{\Lag}{\mathscr{L}}

\def\G{\mathcal{G}}
\def\bt{\beta_\tau}
\def\bp{\beta_\varphi}

\begin{document}




\title{Phase transitions of black strings in dynamical Chern-Simons modified gravity}


\author{Crist\'obal Corral}
\email{crcorral@unap.cl}
\affiliation{Instituto de Ciencias Exactas y Naturales and Facultad de Ciencias, Universidad Arturo Prat, Avenida Arturo Prat Chac\'on 2120, 1110939, Iquique, Chile}

\author{Cristi\'an Erices}
\email{cristian.erices@ucentral.cl}
\affiliation{Universidad Central de Chile, Vicerrector\'ia Acad\'emica, Toesca 1783, Santiago, Chile.}
\affiliation{Universidad Cat\'olica del Maule, Av. San Miguel 3605, Talca, Chile.}

\author{Daniel Flores-Alfonso}
\email{daniel.flores@cinvestav.mx}
\affiliation{Departamento de Física, CINVESTAV-IPN, A.P. 14-740, C.P. 07000, Ciudad de México, Mexico}

\author{Kristiansen Lara}
\email{kristiansen.lara@usach.cl}
\affiliation{Departamento de F\'isica, Universidad de Santiago de Chile, Avenida Ecuador 3493, Santiago, Chile}

\begin{abstract}
We study conserved charges and thermodynamics of analytic rotating anti-de Sitter black holes with extended horizon topology---also known as black strings---in dynamical Chern-Simons modified gravity. The solution is supported by a scalar field with an axionic profile that depends linearly on the coordinate that spans the string. We compute conserved charges by making use of the renormalized boundary stress-energy tensor. Then, by adopting the Noether-Wald formalism, we compute the black string entropy and obtain its area law. Indeed, the reduced Euclidean Hamiltonian approach shows that these methods yield a consistent first law of thermodynamics. Additionally, we derive a Smarr formula using a radial conservation law associated to the scale invariance of the reduced action and obtain a Cardy formula for the black string. A first-order phase transition takes place at a critical temperature between the ground state and the black string, above which the black string is the thermodynamically favored configuration. 
\end{abstract}

\maketitle

\section{Introduction\label{sec:intro}}

Chern--Simons modified gravity (CSMG) is a well-known scalar-tensor theory in four dimensions which was first proposed in Ref.~\cite{Jackiw:2003pm}. It is motivated by anomaly cancellation in curved spacetimes, particle physics, and by the low-energy limit of string theory (for a review see~\cite{Alexander:2009tp}). The theory is endowed with a nonminimal coupling between a scalar field and the Chern-Pontryagin density. Quite notably, this interaction captures parity-violating features in the strong gravity regime. Moreover, under certain conditions (see Ref.~\cite{Ahmedov:2010rn}) its equations of motion effectively reduce to those of topologically massive gravity~\cite{Deser:1981wh,Deser:1982vy}---a 3D gravity theory constructed from a Chern--Simons form. This theory does not belong to the Horndeski family of theories~\cite{Horndeski:1974wa}, since the Chern--Simons coupling generates higher-order field equations. Thus, in order to avoid drawbacks with the Cauchy initial-value problem, it should be considered as an effective theory arising from the UV completion of general relativity (GR)~\cite{Delsate:2014hba}.

Different physical consequences of this theory have been explored since its conception. For instance, the Chern--Simons coupling yields a nontrivial contribution to the post-Newtonian parameter related to frame dragging and gyroscopic precession~\cite{Alexander:2007vt}; observables that could be tested in the near future~\cite{Alexander:2007zg}. When the scalar field is nondynamical and invariant under $SO(3)$, all static and spherically symmetric solutions of GR remain solutions of CSMG since the Pontryagin density vanishes identically; this is not true for axially symmetric configurations, where only slowly rotating and numerical black hole solutions with and without NUT charge are known~\cite{Konno:2007ze,Grumiller:2007rv,Alexander:2007zg,Alexander:2007kv,Konno:2008np,Yunes:2009hc,Cambiaso:2010un,Brihaye:2016lsx}. Shadows of the latter were analyzed in~\cite{Amarilla:2010zq} by integrating null geodesics, showing how the Chern--Simons coupling modifies the shape of the shadow. Perturbations and quasinormal modes around Schwarzschild and slowly rotating black holes have been recently attracted a lot of interest and they have been studied in Refs.~\cite{Bhattacharyya:2018hsj,Okounkova:2019dfo,Okounkova:2019zjf,Wagle:2021tam,Srivastava:2021imr}. On the other hand, CSMG is the only parity-violating extension of GR compatible with luminal propagation of gravitational waves~\cite{Nishizawa:2018srh}. However, the nonminimal coupling does affect the polarization of the latter, producing an amplitude birefringence in their propagation~\cite{Alexander:2017jmt}. Regarding holographic applications, the scalar-Chern-Simons coupling generates spontaneous angular momentum on the conformal field at the boundary, even for static configurations in the bulk~\cite{Liu:2012zm}. Moreover, this theory provides a simple model to study Hall viscocity of holographic fluids; the latter depending only on near-horizon information of the background~\cite{Saremi:2011ab,Chen:2011fs}.    

Recently, analytic rotating solutions were found in CSMG by relaxing the symmetry assumptions on the scalar field~\cite{Cisterna:2018jsx} (see for instance Refs.~\cite{Grumiller:2007rv,Bardoux:2012aw,Cisterna:2017qrb}). These solutions represent rotating black holes with an extended horizon. They correspond to the uplift of the Bañados--Teitelboim--Zanelli (BTZ) black hole~\cite{Banados:1992wn,Banados:1992gq} to four dimensions supported by scalar fields with axionic profile: a homogeneous BTZ black string. Black strings are related to line mass gravitational sources in the same way black holes relate to point mass sources. Homogeneous black strings are product spaces, i.e., a black hole times a one-dimensional space. In this work, we strictly consider homogeneous black strings, in particular those which have locally $\mbox{AdS}_3\times\mathbb{R}$ asymptotics (see~\cite{Gregory:2000gf} for the flat case). These spaces bear some resemblance with Nariai and Bertotti-Robinson spacetimes~\cite{Nariai:1951,Bertotti:1959pf,Robinson:1959ev} and solutions in flux compactification of string theory~\cite{Douglas:2006es}. Indeed, the Chern--Simons modification to GR yields a highly nontrivial contribution to the field equations in presence of rotation, rendering this class of solution scarce. Moreover, when metric and connection are treated as independent variables in the tetrad formulation, BTZ black strings with nontrivial torsion were found in~\cite{Cisterna:2018jsx}.
Thus, these are valuable configurations for the study of conserved charges and thermodynamics of rotating objects in CSMG, as we do here.

Generically, black strings are higher-dimensional asymptotically flat vacuum solutions with extended topology of their horizon that represent interesting counterexamples to topological censorship~\cite{Friedman:1993ty} and to uniqueness theorems in higher-dimensional GR~\cite{Israel:1967za,Carter:1971zc,Wald:1971iw}. They are also known in the presence of nonlinear matter content and in theories beyond GR (for an incomplete list of developments see~\cite{Chan:1994ad,Giribet:2006ec,Astorino:2018dtr,Cisterna:2018jsx,Cisterna:2018mww,Nakas:2019rod,Giacomini:2019qov,Cisterna:2020kde,Hendi:2020apr,Hennigar:2020drx,Arratia:2020hoy,Cisterna:2021ckn,Canfora:2021ttl}). Nevertheless, it is well-known that these solutions usually suffer from the Gregory--Laflamme instability: a long-wavelength linear instability driven by an unstable mode traveling along the extended direction~\cite{Gregory:1993vy,Gregory:1994bj,Gregory:2000gf}. In fact, numerical evolution suggests that this instability remains at the non-linear regime~\cite{Choptuik:2003qd,Lehner:2010pn}, as well as perturbatively in Lovelock theory~\cite{Giacomini:2015dwa,Giacomini:2016ftc}. 
On the other hand, constructing homogeneous black strings in GR with a nonvanishing cosmological constant is a nontrivial task, due to the dynamics. However, this issue can be circumvented by introducing scalar fields with an axionic profile, i.e., with a linear dependence on the extended flat coordinates ~\cite{Cisterna:2018jsx,Cisterna:2017qrb,Cisterna:2018mww}. Interestingly enough, this approach allows one to find black strings in four dimensions, despite the fact that these solutions were first believed to be strictly higher-dimensional objects. Additionally, there exists evidence that axionic fields might stabilize the Gregory--Laflame instability at the perturbative level in four dimensions and they could play an important role in the final state of black strings with axionic charge~\cite{Cisterna:2019scr}; something that it is not guaranteed in higher dimensions~\cite{Henriquez-Baez:2021gdn,Dhumuntarao:2021gdb}.  Nevertheless, a complete proof of stabilization of axionic black strings in four dimensions at the full nonlinear level is still lacking.

In this work, we study conserved charges and thermodynamics of black strings in dynamical Chern--Simons modified gravity sourced by a scalar field with axionic profile. To this end, we first compute conserved charges using the renormalized boundary stress-energy tensor. This allows us to obtain the mass and angular momentum of the solution. Then, we compute the black string entropy via the Noether-Wald formalism and obtain the corresponding area law. Moreover, we carry out the Hamiltonian approach and show explicitly that the conserved charges and the black string entropy satisfy the first law of thermodynamics. The Euclidean on-shell action is computed by adding to it standard boundary terms that render its value finite while defining a well-posed variational principle. To lowest order in the saddle-point approximation, we obtain the free energy, mass, entropy, and angular momentum of the solution, and show that the first law of thermodynamics is satisfied. We also obtain a consistent Smarr formula following the approach of Ref.~\cite{BT}. Additionally, we find the ground state and a non-Einstein product space that belong to the space of solutions. We compute the mass of the former through standard Euclidean methods and, as it might be expected, they turn out to be negative, similar to the hairy solitons found in Refs.~\cite{Correa:2010hf,Correa:2011dt,Gonzalez:2011nz,Anabalon:2016izw,Ayon-Beato:2019kmz}. Remarkably, by analyzing global thermodynamic stability, we find that the black string can develop a phase transition below a critical temperature. Critical phenomena of black objects has become an active area of research since the seminal works of Hawking-Page~\cite{Hawking:1982dh} and it has been extensively studied until today. Just to mention the well-known cases in the literature, in the four dimensional canonical approach, i.e. with fixed cosmological constant, phase transitions of Reissner-N\"{o}rdstrom-AdS and Kerr-Newman-AdS have been analysed. In the former, the black hole free energy exhibits a `cusp' behavior respect to the temperature, which means that there are two branches of black holes, large and small ones \cite{chamblin99}. In the latter, phase transitions between the large and small black holes can disappear for sufficiently large electric charge or angular momentum \cite{caldarelli99}. From a qualitative perspective, the kind of phase transitions that the black string configuration suffers at a critical temperature, resembles the three-dimensional case, where the critical temperature at which thermal AdS collapse to the BTZ black hole, is parametrized by the AdS radius. There is only one branch of BTZ black holes. In our black string, this behavior is similar, but the critical temperature is determined by the axionic charge.

The article is organized as follows. In Sec.~\ref{sec:CSMG} we review Chern--Simons modified gravity as well as the rotating black string solutions with nontrivial scalar field obtained in Ref.~\cite{Cisterna:2018jsx}. In Sec.~\ref{sec:ACBS} we compute the asymptotic charges through the renormalized boundary stress-energy tensors. Then, in Sec.~\ref{sec:HPPT} we compute the black string entropy and the renormalized Euclidean on-shell action to obtain the relevant thermodynamic quantities. We derive the Smarr formula for these extended solutions in Sec.~\ref{sec:smarrBS}. Section~\ref{sec:solitons} is devoted to presenting the ground state and a new non-Einstein conformally flat product space as solutions to CSMG. Both are supported by scalar fields with linear dependence on the flat and time coordinate, respectively. We show explicitly that the mass of the ground state is negative and the non-Einstein conformally flat product space is topologically $\mathbb{R}\times S^3$. In Sec.~\ref{sec:local} we show that, below a certain critical temperature, the ground state is thermodynamically favored over the black string, developing a first order Hawking-Page phase transition. Finally, in Sec.~\ref{sec:conclusions} we present our conclusions and further remarks.


\section{Rotating Black strings in Chern--Simons modified gravity\label{sec:CSMG}}

Let us review the black string solutions found in Ref.~\cite{Cisterna:2018jsx}, as they are the main subject of this work. The theory under consideration is given by the Einstein--Hilbert action modified by the nonminimal coupling between the Chern--Pontryagin density and a dynamical scalar field. This is to say, we focus on the action principle
\begin{align}\label{actionCSMG}
 I &= I_{\rm bulk} + I_{\rm GHY} + I_{\rm ct}\,,
\end{align}
where
\begin{align}\label{Ibulk}
 I_{\rm bulk} &=  \frac{1}{2\kappa}\int_{\mathcal{V}}\diff{^4x}\sqrt{-g}\left(R-2\Lambda + \frac{\alpha\phi}{4}\, ^*RR   -\kappa g^{\mu\nu}\nabla_\mu\phi\nabla_\nu\phi\right) \equiv \int_{\mathcal{V}}\diff{^4x}\sqrt{-g}\;\Lag_{\rm bulk} \,, \\
 \label{IGHY}
 I_{\rm GHY} &= \frac{1}{\kappa}\int_{\partial\mathcal{V}}\diff{^3x}\sqrt{-h}\left(K + \frac{\alpha\phi}{2}\, n_\mu \epsilon^{\mu\nu\lambda\rho}K_{\nu}{}^\sigma\nabla_\lambda K_{\rho\sigma} \right) \equiv \int_{\partial\mathcal{V}}\diff{^3x}\sqrt{-h}\;\Lag_{\rm GHY} ,\\
 \label{Ict}
 I_{\rm ct} &= \int_{\partial\mathcal{V}}\diff{^3x}\sqrt{-h}\left(\zeta_1 + \zeta_2\mathcal{R} + \zeta_3  h^{\mu\nu}\nabla_\mu\phi\nabla_\nu\phi + ... \right) \equiv \int_{\partial\mathcal{V}}\diff{^3x}\sqrt{-h}\;\Lag_{\rm ct} .
\end{align}
Here, the gravitational constant is defined as $\kappa=8\pi G_N$, with $G_N$ being Newton's constant. The nonminimal coupling is measured by the dimensionful parameter $\alpha$ and the Chern--Pontryagin density is defined as
\begin{align}\label{pontryagin}
 ^*RR \equiv \frac{1}{2}\epsilon^{\lambda\rho\sigma\tau} R^{\mu}{}_{\nu\lambda\rho} R^{\nu}{}_{\mu\sigma\tau} = 2\nabla_\mu\left[\epsilon^{\mu\nu\alpha\beta}\Gamma^\sigma_{\nu\lambda}\left( \partial_\alpha \Gamma^\lambda_{\beta\sigma} + \frac{2}{3}\Gamma^\lambda_{\alpha\gamma} \Gamma^\gamma_{\beta\sigma} \right) \right],
\end{align}
with $\epsilon^{\mu\nu\lambda\rho}$ being the Levi-civita tensor. The induced metric on the boundary $\partial\mathcal{V}$ is given by $h_{\mu\nu}=g_{\mu\nu} - n_\mu n_\nu$, where $n^\mu$ is a space-like normal unit vector, and $h=\det h_{\mu\nu}$ is the determinant of the induced metric. Moreover, $K_{\mu\nu} = h^{\rho}{}_\mu h^{\lambda}{}_\nu\nabla_\rho n_\lambda$ denotes the extrinsic curvature with $K$ being its trace. The first boundary term, $I_{\rm GHY}$, defines a well-posed variational principle with Dirichlet boundary conditions~\cite{Grumiller:2008ie}. It is composed by the standard Gibbons--Hawking--York term~\cite{Gibbons:1976ue,York:1972sj} and an additional piece that involves the nonminimal coupling between the scalar field and the Chern--Simons form for the extrinsic curvature at the boundary~\cite{Grumiller:2008ie}. The second boundary term, $I_{\rm ct}$, guarantees that the on-shell action remains finite in spacetimes with AdS asymptotics, provided a particular choice of the $\zeta$-couplings~\cite{Emparan:1999pm,Balasubramanian:1999re};\footnote{In GR, the renormalization prescription used here is equivalent to the Kounterterms proposed in Ref.~\cite{Olea:2006vd} and to early proposals in~\cite{Aros:1999id,Aros:1999kt,Mora:2004kb} when the Weyl tensor vanishes at the AdS boundary (see Ref.~\cite{Anastasiou:2020zwc}).} in this case, $\mathcal{R}^{\mu\nu}_{\lambda\rho}$ is the intrinsic curvature of the boundary, $\mathcal{R}_{\mu\nu} = \mathcal{R}^{\lambda}{}_{\mu\lambda\nu}$, and $\mathcal{R} = h^{\mu\nu}\mathcal{R}_{\mu\nu}$. Notice that the $\zeta$-couplings have different length units, namely $[\zeta_1]=L^{-3}$, $[\zeta_2]=L^{-2}$ and $[\zeta_3]=L$. Hence, the term $I_{\rm ct}$ is constructed solely from intrinsic quantities in the boundary, unaffecting the bulk dynamics. 

The field equations are obtained by performing variations of the action~\eqref{actionCSMG} with respect to the metric and scalar field, giving
\begin{subequations}\label{eom}
 \begin{align}\label{eomg}
 \mathcal{E}_{\mu\nu} &\equiv R_{\mu\nu} - \frac{1}{2}g_{\mu\nu}R + \Lambda g_{\mu\nu} + \alpha C_{\mu\nu} - \kappa  T_{\mu\nu} = 0,\\
 \label{eomp}
 \mathcal{E} &\equiv \Box\phi + \frac{\alpha}{8\kappa}\, ^*RR = 0,
\end{align}
\end{subequations}
respectively. We use the notation $\Box = g^{\mu\nu}\nabla_\mu \nabla_\nu$ and 
\begin{align}
 T_{\mu\nu} &= \nabla_\mu\phi\nabla_\nu\phi - \frac{1}{2}g_{\mu\nu}\nabla_\lambda\phi\nabla^\lambda\phi,\\
 C^{\mu\nu} &= \nabla_\rho\phi\,\epsilon^{\rho\sigma\lambda(\mu}\nabla_\lambda R^{\nu)}{}_\sigma + \nabla_\rho\nabla_\sigma\phi\, ^*R^{\sigma(\mu\nu)\rho},
\end{align}
with $^*R^{\mu}{}_{\nu}{}^{\lambda\rho} = \tfrac{1}{2}\epsilon^{\lambda\rho\sigma\tau}R^{\mu}{}_{\nu\sigma\tau}$. The tensor $C_{\mu\nu}$ is traceless and it satisfies the property~\cite{Jackiw:2003pm,Alexander:2009tp}
\begin{align}\label{nablaC}
 \nabla_\mu C^{\mu\nu} &= -\frac{1}{8}\,^*RR\nabla^\nu\phi. 
\end{align}
We dub $C_{\mu\nu}$ as the $C$-tensor from hereon. Notice that its contribution to the Einstein field equations involves covariant derivatives of the Riemann tensor, yielding third order equations for the metric. Thus, in order to avoid issues with the initial value formulation, Chern--Simons modified gravity should be considered as an effective field theory coming from their ultraviolet completion~\cite{Delsate:2014hba}.

As shown in Ref.~\cite{Cisterna:2018jsx}, the field equations can be solved by assuming the following \emph{Ans\"atze} for the line element and scalar field
\begin{align}\label{BTZBS}
 \diff{s^2} = -N^2(r)f(r)\diff{t^2} + \frac{\diff{r^2}}{f(r)} + r^2\left(N^\varphi(r)\diff{t} + \diff{\varphi} \right)^2 + \diff{z^2} \;\;\;\;\; \mbox{and} \;\;\;\;\; \phi = \phi(z),
\end{align}
respectively. Under this assumption, the Pontryagin density vanishes identically. This is a consequence of the isometry group used in Eq.~\eqref{BTZBS} rather than an imposition to simplify the field equations. Then, the Klein-Gordon equation is directly solved by
\begin{align}\label{phisol}
    \phi = \phi_0 + \lambda z,
\end{align}
where $\lambda$ is an integration constant usually referred to as the axionic charge. Additionally, the shift symmetry in field space, i.e. $\phi\to\phi - \phi_0$, allows one to set $\phi_0=0$ without loss of generality. 

When $N(r)=1$,\footnote{The condition $N(r)=1$ will be justified a posteriori by Hamiltonian analysis (see Eq.~\eqref{sole} below).} and assuming that the solution of the scalar field in Eq.~\eqref{phisol} holds, the components of the field equations for the metric, namely $g^{\mu\rho}\mathcal{E}_{\rho\nu}=\mathcal{E}^\mu_\nu = 0$, become
\begin{align}\notag
\mathcal{E}^t_t &= 2r^3N^{\varphi\,\prime\prime} N^\varphi + r^3 \left(N^{\varphi\,\prime}\right)^2 + 2r\kappa\lambda^2 + 6r^2N^{\varphi\,\prime} N^\varphi + 4r\Lambda  + 2f^{\prime} \\
\notag
&\quad + \lambda\alpha\bigg[ 6r^4 N^{\varphi\, \prime\prime} N^{\varphi\,\prime} N^\varphi  + 2r^4 \left(N^{\varphi\,\prime}\right)^3 + 12r^3 \left(N^{\varphi\,\prime}\right)^2 N^\varphi + 2r^2 N^{\varphi\,\prime\prime\prime} f - r^2 f^{\prime\prime\prime} N^\varphi \\
&\quad+ r^2 N^{\varphi\,\prime\prime}f^{\prime}  - r^2f^{\prime\prime} N^{\varphi\,\prime} + 10rN^{\varphi\,\prime\prime}f + 4rN^{\varphi\,\prime}f^{\prime} + 6N^{\varphi\,\prime}f\bigg] =0\,, \\
\notag
\mathcal{E}^r_r &= r^3\left(N^{\varphi\,\prime}\right)^2 + 2r\kappa\lambda^2 + 4r\Lambda  + 2f^{\prime} +  \lambda\alpha\bigg[ 2r^4\left(N^{\varphi\,\prime}\right)^3 + r^2 N^{\varphi\,\prime\prime}f^{\prime} - r^2f^{\prime\prime} N^{\varphi\,\prime} \\
&\quad - 2rN^{\varphi\,\prime\prime}f + 4rN^{\varphi\,\prime}f^{\prime} - 6N^{\varphi\,\prime}f\bigg] =0 \,,\\
\notag
\mathcal{E}^\varphi_\varphi &= 2r^3 N^{\varphi\,\prime\prime} N^\varphi + 3r^3\left(N^{\varphi\,\prime}\right)^2 - 2r\kappa\lambda^2 + 6r^2N^{\varphi\,\prime} N^\varphi - 4r\Lambda - 2rf^{\prime\prime} \\
\notag
&\quad +  \lambda\alpha\bigg[6r^4N^{\varphi\,\prime\prime} N^{\varphi\,\prime} N^\varphi + 4r^4\left(N^{\varphi\,\prime}\right)^3 + 12r^3\left(N^{\varphi\,\prime}\right)^2N^\varphi + 2r^2N^{\varphi\,\prime\prime\prime}f   \\
&\quad - r^2f^{\prime\prime\prime} N^\varphi + 2r^2N^{\varphi\,\prime\prime}f^{\prime} - 2r^2f^{\prime\prime} N^{\varphi\,\prime} + 8rN^{\varphi\,\prime\prime}f + 8rN^{\varphi\,\prime}f^{\prime}\bigg] =0 \,, \\
\mathcal{E}^t_\varphi &= 2r^3N^{\varphi\,\prime\prime} + 6r^2N^{\varphi\,\prime} + \lambda\alpha\left[6r^4N^{\varphi\,\prime\prime} N^{\varphi\,\prime} + 12r^3\left(N^{\varphi\,\prime}\right)^2 - r^2f^{\prime\prime\prime} \right] =0 \,,\\
\mathcal{E}^z_z &= - r^3 \left(N^{\varphi\,\prime}\right)^2 - 2r\kappa\lambda^2 + 4\Lambda r + 2rf^{\prime\prime} + 4f^{\prime}=0  \,,
\end{align}
where prime denotes differentiation with respect to the radial coordinate $r$. Additionally, since the trace of the C-tensor vanishes, we have
\begin{align}
    g^{\mu\nu}\mathcal{E}_{\mu\nu} &= - r^3 \left(N^{\varphi\,\prime}\right)^2 + 2r\kappa\lambda^2 + 8\Lambda r + 2rf^{\prime\prime}+ 4f^{\prime} = 0\,.
\end{align}
The equation $\mathcal{E}^z_z - g^{\mu\nu}\mathcal{E}_{\mu\nu}=0$ implies immediately that the integration constant associated to the scalar field is fixed in terms of the cosmological constant through $\lambda^2 = - \Lambda/\kappa$. Thus, in order for the scalar field to be real, the condition $\Lambda<0$ must be met in order to avoid problems with unitarity. Additionally, the equation
\begin{align}
    \mathcal{E}^t_t - \mathcal{E}^r_r - N^\varphi \mathcal{E}^t_\varphi = 2\lambda\alpha f\left[r^2 N^{\varphi\,\prime\prime\prime} + 6rN^{\varphi\,\prime\prime} + 6N^{\varphi\,\prime} \right] = 0\,,
\end{align}
implies that, for $\lambda\alpha \neq0$, the metric function $N^\varphi(r)$ can be integrated as
\begin{align}\label{Nsol}
 N^\varphi(r) = j_0 + \frac{j_1}{r} + \frac{j}{2r^2}.     
\end{align}
Indeed, the condition $\lambda\alpha \neq0$ shows that the rotating solution is supported only if the axionic charge or Chern-Simons coupling $\alpha$ are nonvanishing. It is worth mentioning that, in general relativity, the asymptotically (A)dS cylindrical black hole of Ref.~\cite{Lemos:1994xp} does not require any scalar field. This is related to its horizon's topology, $\mathbb{T}^2$; in such a case, the equations of motion does not impose any constraints on the cosmological constant whatsoever. Nevertheless, the topology of the homogeneous black string's horizon studied here, $\mathbb{S}^1\times\mathbb{R}$, requires the existence of a nontrivial scalar field $\phi=\lambda z$ alongside the condition $\lambda\alpha \neq0$, due to the presence of the scalar-Chern-Simons coupling.

Replacing Eq.~\eqref{Nsol} into the remaining components of the field equations, one finds that $j_1=0$ and the metric function $f(r)$ is found to be 
\begin{align}\label{fsol}
 f(r) = -m + \frac{r^2}{\ell^2} + \frac{j^2}{4r^2} \;\;\;\;\; \mbox{with} \;\;\;\;\; \ell^{-2}=-\frac{\Lambda}{2}\,. 
\end{align}
The line element~\eqref{BTZBS}, alongside the metric functions~\eqref{Nsol} and~\eqref{fsol} ($j_1=0$), illustrates the rotating black string found in Ref.~\cite{Cisterna:2018jsx}. It corresponds to the uplifting of the BTZ black hole~\cite{Banados:1992wn,Banados:1992gq} with a nontrivial scalar field into Chern-Simons modified gravity. It is locally equivalent to $\mbox{AdS}_3\times\mathbb{R}$, cf.~\cite{Ayon-Beato:2004ehj}. Its horizons are determined by the condition $f(r_\pm)=0$ and are located at
\begin{align}\label{horizons}
 r_\pm = \ell \left[\frac{m \pm \sqrt{m^2 - j^2/\ell^2}}{2} \right]^{1/2}.
\end{align}
Thus, for their existence, the condition $m^2 - j^2/\ell^2\geq0$ must be met. The extremal case is obtained when the latter bound is saturated. As in the BTZ black hole, all curvature invariants are constant and depend solely on the AdS radius. However, identifications of points of AdS spacetime by a discrete subgroup of $SO(2,2)$ yields a singularity in the causal structure where closed time-like curves arise~\cite{Banados:1992gq}. The same type of singularity appears in the BTZ black string above. Finally, notice that the AdS radius is neither $\ell^{-2}=-\Lambda$ nor $\ell^{-2}=-\Lambda/3$, as in dimensions $D=3$ and $D=4$, respectively. This is due to the presence of axionic fields that generate an effective cosmological constant.

\section{Conserved charges\label{sec:ACBS}}

Identifying the conserved charges of a gravitational configuration is an important issue in standard gravity. In extensions to it, the issue becomes even more cumbersome. There have been many techniques developed for this purpose, each approach has its benefits and drawbacks. In Chern-Simons modified gravity, conserved charges have been studied in the context of background Killing vector symmetries~\cite{Tekin:2007rn} and quasilocal methods~\cite{Liu:2012zm}. In this section, we use the latter to compute the mass and angular momentum of the black string. Intuitively, since the transverse section of the black string are BTZ black holes, we expect the mass and angular momentum per unit length of the black string to be related to these quantities. In what follows we sort out the specifics of the computations in CSMG.

First, it is well-known that, in spacetimes with AdS asymptotics, the standard Brown--York stress tensor~\cite{Brown:1992br} needs to be suplemented by boundary counterterms to render conserved charges finite~\cite{Balasubramanian:1999re}. The boundary stress-energy tensor of Chern--Simons modified gravity is 
\begin{align}\notag
    T^{\rm(bdy)}_{\mu\nu} &= -\frac{1}{2\kappa}\left(2K_{\mu\nu} - 2h_{\mu\nu}K + T^{\rm (CS)}_{\mu\nu} + T^{\rm(GMM)}_{\mu\nu} \right)  \\ 
    \label{Tbdy}
   &\quad   + \zeta_1 h_{\mu\nu} - 2\zeta_2\left(\mathcal{R}_{\mu\nu} - \frac{1}{2}h_{\mu\nu}\mathcal{R} \right) - 2\zeta_3 T^{(\phi)}_{\mu\nu} \,,
\end{align}
where $T^{\rm (CS)}_{\mu\nu}$ arises from the variation with respect to the induced metric of the scalar-Chern--Simons term (for instance see~\cite{Liu:2012zm}). On the other hand, and $T^{\rm(GMM)}_{\mu\nu}$ appears from the second piece of Eq.~\eqref{IGHY}, proposed by Grumiller, Mann, and McNees (GMM) in Ref.~\cite{Grumiller:2008ie}. However, it is known that in asymptotically AdS spacetimes these two contributions decay fast enough towards the boundary and they do not contribute to the quasilocal charges~\cite{Liu:2012zm}. Additionally, the last four terms of Eq.~\eqref{Tbdy} renormalize the stress-energy tensor in asymptotically AdS spacetimes and they arise from variations of $I_{\rm ct}$ with respect to the induced metric~\cite{Balasubramanian:1999re}. Here, $T^{(\phi)}_{\mu\nu}$ denotes the energy-momentum tensor of the scalar fields constructed out of the induced metric and we do not include the other contributions in $I_{\rm ct}$ since they decay sufficiently fast as $r\to\infty$ (see also~\cite{Caldarelli:2016nni,Flores-Alfonso:2019aae,Flores-Alfonso:2020ayc}). 

The quasilocal charge associated to a Killing vector field $\xi=\xi^\mu\partial_\mu$ is defined through~\cite{Balasubramanian:1999re}
\begin{align}\label{Q}
    Q\left[\xi \right] = \int_\Sigma\diff{^2x}\sqrt{\sigma}\; T^{\rm(bdy)}_{\mu\nu}u^\mu\xi^\nu\,,
\end{align}
where $u^\mu$ is the vector that generates the flow of time in $\partial V$ and $\sigma$ is the determinant of the induced metric on $\Sigma$ (see~\cite{Balasubramanian:1999re} for details). Renormalization of the boundary stress-energy tensor~\eqref{Tbdy} on asymptotically $\mbox{AdS}_3\times\mathbb{R}$ spacetimes demands that the counterterms must satisfy the relation
\begin{align}\label{CTBS}
 \zeta_1 &= - \frac{2\zeta_3 + \ell}{\kappa\ell^2}.
\end{align}
This value can be obtained by expanding~\eqref{Q} at large but fixed radius while choosing $\zeta_1$ such that the divergent terms vanish. Afterward, one can take the limit $r\to\infty$ safely and obtain a finite asymptotic charge.
 
The mass, angular, and linear momentum per unit of length $L$ of the solution are obtained from the Killing vectors field that generate the temporal, rotational, and translational symmetry, respectively; these are
\begin{align}\label{MBS}
    Q\left[\partial_t \right] &\equiv M = \frac{mL}{8G}, \\
    \label{JBS}
        Q\left[\partial_\varphi \right] &\equiv J = \frac{jL}{8G},\\
        Q\left[\partial_z \right] &\equiv P_z = 0\,.
\end{align}
Thus, this method shows that $m$, and $j$, are integration constants related to the mass and angular momentum of the solution. Moreover, homogeneity of the black string along the $z$-direction yields a vanishing conserved charge. Thus, as long as the axionic charge remains fixed in terms of the cosmological constant as $\lambda^2=-\Lambda/\kappa$, it cannot be associated to a conserved quantity. This could provide a first glimpse of a counterexample to the stronger form of correlated stability; at least in higher-dimensions~\cite{Dhumuntarao:2021gdb}. Nevertheless, it is worth mentioning that there exist different approaches where the cosmological constant arises as an integration constant~\cite{Buchmuller:1988wx,Unruh:1988in,Henneaux:1989zc,Ng:1990xz}. In this case, the axionic charge can be interpreted truly as a conserved quantity associated to volume preserving diffeomorphisms, providing a natural scenario to introduce a thermodynamic pressure in the extended phase space approach~\cite{Kubiznak:2014zwa}.

\section{Black string thermodynamics\label{sec:HPPT}}

Our current understanding of black hole thermodynamics is a deep-seated consequence of semiclassical gravity~\cite{Hawking:1975vcx,Wald:1975kc}. It is the means by which we know horizon surface area is not analogous to thermodynamic entropy, rather, it is the black hole's entropy~\cite{Bekenstein:1973ur}. This has been proved by Noether charge analysis~\cite{Wald:1993nt} and through Euclidean methods~\cite{Gibbons:1976ue}. Indeed, many advances have been made possible thanks to the latter approach taken from quantum physics~\cite{GibbonsHawking}. Moreover, black hole thermodynamics has been the inspiration of a plethora of works in theoretical physics and is a key sector of quantum gravity.  Even the simplest black hole distinguishes itself from ordinary thermodynamic substances, such as quantum fluids and ferromagnetic matter~\cite{Callen}. This is due to the degree of homogeneity of a black hole's fundamental relation. Such a number can be read off from a system's Gibbs--Duhem equation, which highlights the significance of a black hole’s Smarr formula~\cite{Smarr:1972kt} (see for example~\cite{Bravetti:2017peq}).

In this section, we obtain the black string entropy using the Noether-Wald formalism. Then, we consider the analytic continuation of the BTZ black string to calculate the Euclidean on-shell action, which corresponds to the free energy of the system. 
The partition function is identified with the Euclidean path integral at first order in the saddle-point approximation around the classical solution~\cite{GH}. Fixing the spacetime boundary conditions is akin to choosing a thermodynamic ensemble. Indeed, in AdS, the boundary conditions for minimally coupled scalar fields with radial profile have been discussed in Ref.~\cite{Henneaux:2006hk}. For homogeneous black strings, the situation is different since the asymptotic behavior is $\mbox{AdS}_3\times\mathbb{R}$ rather than $\mbox{AdS}_4$. Nevertheless, since the axion generates only an effective AdS radius in the transverse section, the standard Henneaux-Teitelboim boundary conditions for AdS gravity~\cite{Henneaux:1985tv} should be enough to ensure well-defined asymptotic charges in this particular case. Still, one should consider all infilling geometries in the path integral and sum over all possible topologies. This is only sensible when enormous symmetry restrictions are enforced, as we consider here. This strategy allows for the thermodynamics of black holes to be completely determined via the Euclidean approach. Then, we employ the reduced Hamiltonian formalism on the Euclidean versions of the black string to calculate its conserved charges. By construction, these charges satisfy the first law of thermodynamics. Indeed, we find that conserved charges and entropy obtained through the Hamiltonian approach coincide with those found in previous sections. Additionally, we derive a Smarr formula using the scale invariance of the reduced action following Ref.~\cite{BT}.

\subsection{Wald entropy}

A useful and covariant method to compute the black string entropy is the Noether-Wald formalism~\cite{Wald:1993nt,Iyer:1994ys}. This is based on the diffeomorphism invariance of a gravitational action principle, generated by a vector field $\xi=\xi^\mu\partial_\mu$, which yields an on-shell conserved current, $J^\mu$, satisfying $\nabla_\mu J^\mu = 0$. The Poincar\'e lemma, in turn, implies that the latter can be written locally as $J^\mu = \nabla_\nu q^{\mu\nu}$, where   
\begin{align}\label{Noetherprep}
    q^{\mu\nu} = -2\left(E^{\mu\nu}_{\lambda\rho}\nabla^\lambda\xi^\rho + 2\xi^\lambda\nabla^\rho E^{\mu\nu}_{\lambda\rho} \right) = -q^{\nu\mu}\,,
\end{align}
is known as the Noether prepotential. Here, $E^{\mu\nu}_{\lambda\rho}$ stands for the functional derivative of the bulk Lagrangian with respect to the Riemann tensor and derivatives thereof (see~\cite{Iyer:1994ys} for details). In particular, for the action in Eq.~\eqref{actionCSMG}, we obtain
\begin{align}
    E^{\mu\nu}_{\lambda\rho} \equiv \frac{\partial\Lag_{\rm bulk}}{\partial R^{\lambda\rho}_{\mu\nu}} = \frac{1}{8\kappa}\left[2\delta_{\lambda\rho}^{\mu\nu} -\, \alpha\phi\left( ^*R^{\mu\nu}{}_{\lambda\rho}+\,^*R_{\lambda\rho}{}^{\mu\nu}\right) \right] \,,
\end{align}
where $\delta^{\mu_1\ldots\mu_p}_{\nu_1\ldots\nu_p}=p!\delta_{[\nu_1}^{[\mu_1}\ldots\delta_{\nu_p]}^{\mu_p]}$ is the generalized Kronecker delta. Conserved charges associated to Killing vector fields $\xi$ are obtained by integrating the Noether prepotential on a codimension-2 hypersurface, $\Sigma$, namely,
\begin{align}\label{NWcharge}
    \mathcal{Q}\left[\xi \right] = \frac{1}{2}\int_{\Sigma}\epsilon_{\mu\nu\lambda\rho}q^{\mu\nu}\diff{x^\lambda}\wedge\diff{x^\rho} \equiv \int_\Sigma \mathcal{Q}_{\mu\nu}\diff{x^\mu}\wedge\diff{x^\nu}\,,
\end{align}
where $\wedge$ is the wedge product of differential forms. Wald showed that the black hole entropy is the Noether charge~\eqref{NWcharge} evaluated at the horizon $\mathcal{H}$~\cite{Wald:1993nt}. More generally, it is obtained by integrating the Noether prepotential over all codimension-2 hypersurfaces that produce obstructions to foliation by functions that define the unitary Hamiltonian evolution of the system~\cite{Hawking:1998jf,Garfinkle:2000ms,Ciambelli:2020qny}.

Since the black string in CSMG is stationary, we consider the isometry generated by the asymptotically time-like Killing vector field that vanishes at the horizon, that is,
\begin{align}\label{Killing}
    \xi &= \partial_t + \Omega\partial_\varphi \,.
\end{align}
Here, $\Omega$ denotes the angular velocity at the horizon---see Eq.~\eqref{Omega&Psi} below.  The relevant components of the dual Noether prepotential for computing the entropy are
\begin{align}
    \mathcal{Q}_{\varphi z} = \frac{r}{4\kappa}\left[ 2f^{\prime} - \alpha\lambda f\left(rN^{\varphi\,\prime\prime}+3N^{\varphi\,\prime} \right)  \right],
\end{align}
where $\phi=\lambda z$ has been used. In the case of the black string, the horizon $\mathcal{H}$ is defined as the codimension-2 hypersurface at constant $t-r$ coordinates, where the metric function $f(r)$ vanishes, i.e. $f(r_+)=0$. Direct integration leads to
\begin{align}\label{SBS}
    S =  \beta_\tau\int_0^{2\pi}\diff{\phi}\int_0^L\diff{z}\,\mathcal{Q}_{\varphi z} \bigg|_{r=r_+} = \frac{\pi r_+ L}{2G} = \frac{\mathcal{A}}{4G}\,,
\end{align}
where $\mathcal{A}=2\pi r_+ L$ is the area of the horizon and $\beta_\tau$ is related to the surface gravity $\mathscr{K}$ through $\beta_\tau^{-1} = \mathscr{K}/(2\pi)$. Indeed, $T=\beta_\tau^{-1}$ is the Hawking temperature of the black string. Thus, the system studied above satisfies the black hole area law, which is an unexpected result in gravity theories beyond GR (for instance see~\cite{Jacobson:1993xs}). 


\subsection{Hamiltonian approach\label{sec:hamiltonian}}

 For the purposes of this work, it is sufficient to consider the analytic continuation of the black hole metric~\eqref{BTZBS} with a base manifold of topology $S^1\times\mathbb{R}$, namely,
\begin{equation}
 \diff{s_E^{2}}=N^2(r) f(r) \diff \tau^{2}+\frac{\diff{} r^{2}}{f(r)}+r^{2}\left(N^\varphi(r) \diff \tau+\diff \varphi\right)^{2}+\diff z^2\,,
\end{equation}
where the Euclidean time range is given by $0\leq \tau\leq \beta_{\tau}$, the angular coordinate satisfies $0\leq \varphi\leq \beta_{\varphi}$ and the flat coordinate goes as $0\leq z\leq L$. Additionally, we consider a scalar field with axionic profile, i.e., $\phi=\phi(z)$. Since the black string has an event horizon, its Euclidean version possesses a degenerate submanifold in its place, called a bolt. The absence of conical singularities at the bolt demands that the period of the Euclidean time, $\beta_\tau$, and the period of the angular coordinate, $\beta_\varphi$ are fixed according to the relations $N(r_+)f'(r_+)=4\pi/\beta_\tau$ and $\beta_\varphi=-\beta_\tau N^\varphi(r_+)$, respectively. Moreover, these quantities are related to the Hawking temperature and angular velocity of the black string at the horizon as $T=\beta_\tau^{-1}$ and $\Omega = -\beta_\varphi/\beta_\tau$, respectively.

The Euclidean action is related to the free energy $\G$ by $I_E=\beta_\tau \G$. Then, the reduced action in Hamiltonian form is 
\begin{equation}\label{reducedaction}
I_E=\beta_\tau\beta_\varphi\int(N\mathcal{H}+N^\varphi\mathcal{H}_{\varphi})\diff r \diff z+B_E\ .
\end{equation}
Here, $B_E$ denotes a boundary term whose variation cancels out all the contributions coming from variations of the bulk action, defining a well-posed variation principle. Additionally, the constraints and the nonvanishing components of the conjugated momenta are respectively given by
\begin{align}
\mathcal{H}&=\frac{r \Lambda}{\kappa}+\frac{f^{\prime}}{2 \kappa}-4 \kappa r\left(\pi^{r\varphi}\right)^{2}+\frac{1}{2} r\left(\frac{\diff{\phi}}{\diff{z}}\right)^{2}\ ,\\
\mathcal{H}_{\varphi}&=-2\, \left(r^{2} \pi^{r \varphi}\right)^{\prime}\ , \\
\pi^{r \varphi}&=\frac{r N^{\varphi\,\prime}}{4 \kappa N}\ .
\end{align}

Performing stationary variations of the action with respect to $\{N,f,N^\varphi,\pi^{r\varphi},\phi\}$, we obtain a set of differential equations equivalent to the ones of Chern-Simons modified gravity in Euclidean signature [see Eq.~\eqref{eom}]. These equations can be solved analytically, recovering the black string solution of Sec.~\ref{sec:CSMG}, namely,
\begin{equation}\label{sole}
\begin{aligned}
N(r)&=1\ ,&\phi(z)&=\lambda z\ , &\pi^{r\varphi}(r)&=-\frac{j}{4\kappa r^2}\ ,\\ N^\varphi(r)&=j_0+\frac{j}{2r^2}\ ,& f(r)&=\frac{r^2}{\ell^2}-m-\frac{j^2}{4r^2}\ , &\frac{1}{\ell^2}&=\frac{\kappa\lambda^2}{2}\ ,
\end{aligned}
\end{equation}
where $\lambda, m, j_0$ and $j$ are integration constants. Moreover, the scalar field is rigid since $\lambda$ is a constant without variation, fixed in terms of the cosmological constant. It is worth mentioning that, although the scalar field is endowed with a linear dependence on the $z$ coordinate, the shift symmetry protects the theory for any stability problem related to the unboundedness of the $z$ coordinate. This can be seen by noticing that all physical quantities depends purely on derivatives of the scalar field. On the other hand, one could impose periodicity on the $z$ coordinate such that it becomes compact; in that case, the topology of the horizon would be $\mathbb{S}^1\times\mathbb{S}^1$, representing a black ring rather than a black string.

The variation of the boundary term, on the other hand, yields
\begin{align}
\delta B_E&= -\bt\bp L \left[\frac{N\delta f}{2\kappa} - 2r^2N^\varphi\delta\pi^{r\varphi}\right]^{r=\infty}_{r=r_+}  \,.
\end{align}
To compute the latter, we first evaluate the variation of the boundary term on the solution given in Eq.~\eqref{sole}. Then, taking into account that the variation of the fields at the bolt are
 \begin{align}
     \delta f|_{r_+}=-\frac{4\pi}{\bt}\delta r_+\;\;\;\;\; \mbox{and} \;\;\;\;\; \delta \pi^{r\varphi}|_{r_+}=-\frac{\delta j}{4\kappa r_+^2}\ ,
 \end{align}
then it follows that the variation of the boundary term at the bolt and infinity are respectively given by
 \begin{align}
 \delta B_E(\infty)&= \frac{\bt\bp L}{2\kappa} \left[\delta{m} - j_0\delta j\right]\, , \\
     \delta B_E(r_+)&= \frac{\bt\bp L}{2\kappa}\left[\frac{4\pi\delta r_+}{\bt }-\left(j_0+\frac{j}{2r_+^2}\right)\delta j\right]\, . 
\end{align}
Thus, we conclude that the total variation of the boundary term, evaluated at the solution~\eqref{sole}, becomes
\begin{equation}
    \delta B_E=\delta B_E(\infty)-\delta B_E(r_+)=\bt\delta\left(\frac{\sigma m}{2\kappa}\right)-\frac{2\pi}{\kappa}\delta A_+ + \bt\frac{j}{2r_+^2}\delta\left(\frac{\sigma j}{2\kappa}\right)\ ,
\end{equation}
where $\sigma=\bp L$ stands for the volume of the base manifold and $A_+=\sigma r_+$ for the area of the black string horizon. From this expression, we identify
 \begin{align}\label{Omega&Psi}
     \Omega=\frac{j}{2r_+^2}\, ,
 \end{align}
as the conjugated thermodynamic variable associated to the angular momentum: the angular velocity of the black string. 
 
Hereafter, we work in the grand canonical ensemble, where the temperature and the angular velocity are kept fixed at the bolt. This allows us to integrate the variation of the boundary term. Its value determines the reduced on-shell action, giving
 \begin{equation}
 I_E=B_E=\bt\left(\frac{\sigma m}{2\kappa}\right)-\frac{2\pi}{\kappa} A_+ -\bt\Omega\left(-\frac{\sigma j}{2\kappa}\right)\ ,
 \end{equation}
 up to an arbitrary additive constant without variation. Since the Gibbs free energy and the Euclidean action are related by $I_E=\bt \G=\bt M-S-\bt \Omega J$, the conserved charges and entropy can be obtained through the standard thermodynamical relations
 \begin{subequations}\label{charges}
  \begin{align}
 M&=\left(\frac{\partial}{\partial \bt}-\bt^{-1}\Omega\frac{\partial}{\partial\Omega}\right)I_E=\frac{\sigma m}{2\kappa}\ ,\\
 J&=-\frac{1}{\bt}\frac{\partial I_E}{\partial \Omega}=-\frac{\sigma j}{2\kappa}\ ,\\
 S&=\left(\bt\frac{\partial}{\partial\bt}-1\right)I_E=\frac{A_+}{4G}\ .
 \end{align}
 \end{subequations}
 It is reassuring to verify that the Noether-Wald formalism and the Hamiltonian approach lead to the Bekenstein-Hawking entropy, considering that non-minimally coupled scalar fields can produce additional contributions such as in other scalar-tensor theories~\cite{Ashtekar:2003jh,Barlow:2005yd,Martinez:2005di}. The sign flip in the angular momentum is a consequence of the computation in Euclidean signature. By construction, the first law of black hole thermodynamics is satisfied, namely, 
 \begin{align}
 \delta M=T\delta S+\Omega\delta J\,.    
 \end{align}
Notice that $j_0$ plays no role in the thermodynamics. This is just a reflection of the fact that we can consider a rotating asymptotic geometry $\mbox{AdS}_3\times\mathbb{R}$ where $j_0$ emerges as the boost parameter along the $t-\varphi$ plane in this region. In consequence, since the angular momentum of the black string is measured respect to this background, without loss of generality, we can always choose the comoving frame where $j_0=0$. Therefore, we conclude that the Hamiltonian approach leads to the same conserved quantities as the renormalized boundary stress tensor. 

\subsection{Smarr formula\label{sec:smarrBS}}

This section is devoted to deriving a Smarr formula for the black string using a Euclidean Hamiltonian approach, as in Ref.~\cite{BT}. To this end, we compute a radial conservation law associated to the scale invariance of the reduced action. Let us recall that only regular solutions may dominate the gravitational path integral. Thus, we impose that matter fields must be finite at the bolt and additionally that they vanish at infinity, which is satisfied along a finite section of the base manifold.

Let us consider a minisuperspace determined by the metric~\eqref{BTZBS}. Then, the reduced action of Chern-Simons modified gravity [see Eq.~\eqref{reducedaction}] is invariant under the set of transformations $\Phi(x)\to\bar{\Phi}(\bar{x})$ generated by the rescaling along the radial coordinate $\bar{r}=\delta r$, where $\delta$ is a positive constant, $\Phi=\left\{N,f,N^\varphi,\pi^{r\varphi},\phi\right\}$, and
\begin{equation}\label{trans}
\begin{aligned}
\bar{N}(\bar{r})&=\delta^{-2} N(r)\ ,&\bar{f}(\bar{r})&=\delta^{2} f(r)\ ,&\bar{\pi}^{r\varphi}(\bar{r})&=\pi^{r \varphi}(r)\ ,\\
\bar{\phi}(z)&=\phi(z)\ ,&\bar{N}^{\varphi}(\bar{r})&=\delta^{-2} N^\varphi(r)\ .
\end{aligned}
\end{equation}
The Noether theorem, in turn, implies that there exists a conserved quantity $C(r)$ associated to radial rescalings, such that $C^{\,\prime}(r)=0$. Using the constraint $\mathcal{H}=0$, we rewrite the conserved quantity as
\begin{equation}
C(r)=\left[N\left(\frac{f'}{2\kappa}r-\frac{f}{\kappa}\right)+N^{\varphi} (4 r^{2} \pi^{r \varphi})\right]\sigma\ .
\end{equation}
Since the latter represents a conserved quantity along the radial direction, it must satisfy $C(\infty)=C(r_+)$. A direct computation using the black string solution~\eqref{BTZBS} with \eqref{phisol}, \eqref{Nsol} and \eqref{fsol}, yields
\begin{align}
C(r_+)=\left(r_+f'(r_+)+\frac{j^2}{r_+^2}\right)\frac{\sigma}{2\kappa}=T S+2 \Omega J\;\;\;\;\; \mbox{and} \;\;\;\;\; C(\infty)=\frac{m \sigma}{\kappa}=2 M\ ,
\end{align}
where the definition of thermodynamic variables in Sec.~\ref{sec:hamiltonian} have been used. Thus, by comparing these two expressions, we obtain the Smarr relation for the black string, namely, 
\begin{equation}\label{smarr}
M=\frac{1}{2}T S+\Omega J\ .
\end{equation}
Morover, the very same relation can be obtained by direct application of the Euler theorem for homogeneous functions, since the entropy can be regarded as a homogeneous function of degree $1/2$ in $(M,J)$, i.e. $ S(\lambda M,\lambda J)=\lambda^{1/2}S(M,J)$. Then,
\begin{equation}\label{r1}
\frac{1}{2} S=\left.M \frac{\partial S}{\partial M}\right|_{J}+\left.J \frac{\partial S}{\partial J}\right|_{M}\ .
\end{equation}
Using the first law $T\delta S=\delta M-\Omega\delta J$ we obtain,
\begin{equation}\label{r2}
\left.\frac{\partial S}{\partial M}\right|_{J}=\frac{1}{T}\ , \quad \left.\frac{\partial S}{\partial J}\right|_{M}=-\frac{\Omega}{T}\ .
\end{equation}
It is straightforward to show that Eqs.~\eqref{r1} and~\eqref{r2} lead to~\eqref{smarr}. This relation is the same as the Smarr formula of a rotating BTZ black hole in three dimensional gravity~\cite{erices1}. Of course, the Smarr relation is independent of the Euclidean or Lorentzian signature.

\section{Ground state and conformally flat non-Einstein product spaces\label{sec:solitons}}

The conformal nature of quantum gravity on AdS$_3$~\cite{Brown:1986nw} and the Cardy formula for the asymptotic growth of the number states in 2D conformal field theories~\cite{Cardy:1986ie} has allowed for the microscopic derivation of the BTZ black hole entropy~\cite{Strominger:1997eq}. This established AdS$_3$ as the ground state of the theory. In higher dimensions, the AdS soliton was later found to play the same role~\cite{Horowitz:1998ha,Galloway:2001uv}. Currently, Cardy formulae are written so as to relate the entropy of a black hole with its mass and the mass of its associated ground state even for more general asymptotic behavior, see for instance~\cite{Correa:2010hf,Correa:2011dt,Gonzalez:2011nz}. This motivates us to establish the ground state of the BTZ black string under consideration.

By noticing that the homogeneous foliation of the four dimensional metric~\eqref{BTZBS} admits an analytic continuation along the symmetries generated by its Killing vector fields $\partial_t$ and $\partial_\varphi$, we propose an \emph{Ansatz} of the form
\begin{equation}
    \diff{s^2} = -N^2(r)f(r)\diff{t^2} + \frac{\diff{r^2}}{f(r)} + h(r)r^2\diff{\varphi^2} + \diff{z^2}\,. \label{soliton2}
\end{equation}

The metric~\eqref{soliton2} solves the field equations of Chern-Simons modified gravity, recall Eq.~\eqref{eom}, when
\begin{align}\label{solsoliton2}
     f(r) &= -\mu + \frac{r^2}{\ell^2}\,, & N^2(r) &= \frac{r^2}{f(r)\ell^2}\,, & h(r) &= \frac{f(r)\ell^2}{r^2}\,, & \phi(z) &= \lambda\,z\,,
\end{align}
where $\ell^{-2} = -\Lambda/2$ is the AdS radius, $\mu$ is an integration constant, and the axionic charge $\lambda$ is fixed in terms of the cosmological constant through $\lambda^2=-\Lambda/\kappa$. 
Reality of the scalar field, on the other hand, implies that the cosmological constant must be negative. Positivity of the Riemannian section of~\eqref{soliton2} demands that the radial coordinate must be restricted to $r\geq r_s\equiv\ell\sqrt{\mu}$. The absence of conical singularities imposes that $\mu>0$ and $\varphi\sim\varphi+\beta_\varphi^{(s)}$, where $\beta^{(s)}_\varphi= \tfrac{4\pi}{f'(r_s)\ell}=\tfrac{2\pi}{\sqrt{\mu}}$. With these conditions, the line element~\eqref{soliton2} with metric functions~\eqref{solsoliton2} represents the ground state with topology $\mbox{AdS}_3\times\mathbb{R}$.

Its mass can be computed directly through standard Euclidean methods of Sec.~\ref{sec:HPPT}. To do so, we consider the Gibbs-Duhem relation in absence of entropy. Moreover, the temperature is arbitrary due to their lack of horizon.  Thus, the mass of the ground state~\eqref{solsoliton2} is given by 
\begin{equation}
    M_0 =-\frac{L}{8 G}\,.
    \label{M2}
\end{equation}
It is clear that the mass of the ground state in Eq.~\eqref{M2} is negative. The negative-energy contribution can be interpreted as the Casimir energy that is generated in the dual field theory when the fermions are antiperiodic on the compact coordinate~\cite{Coussaert:1993jp}.
Additionally, it is direct to see that all curvature invariants associated to this solution remain finite. Therefore, it represents a negative-energy regular solution without horizon. These configuration are relevant when studying phase transitions, as we show further below. 

The Casimir energy interpretation stems from the Cardy formula. Before writing it here, let us remark that we have used in Eq.~\eqref{solsoliton2} the same notation as for the black string. In the Cardy formula, states of the same configuration are compared. Thus, the ground state and the black string must be comparable, i.e., they have the same boundary conditions and global symmetries. Consequently, the entropy of the black string can be written as
\begin{equation}
S=4\pi l\left( \sqrt{\frac{-M_0\Delta^+}{2}} + \sqrt{\frac{-M_0\Delta^-}{2}}\right),
\end{equation}
where $\Delta^{\pm}=(M\pm J/l)/2$, cf. \cite{BravoGaete:2017dso}. This is the Cardy formula for the analytic rotating black string presented in Sec.~\ref{sec:CSMG}.

Interestingly enough, metric~\eqref{BTZBS} admits another analytic continuation; this one along the symmetries generated by the Killing fields $\partial_t$ and $\partial_z$. Hence, we now explore a metric \emph{Ansatz} of the form
\begin{equation}
    \diff{s^2} = - \diff{t^2} + \frac{\diff{\rho^2}}{u(\rho)} + \rho^2\left(v(\rho)\diff{z} + \diff{\varphi} \right)^2 + u(\rho)\diff{z^2}\,, 
    \label{soliton1}
\end{equation}
which solves the field equations~\eqref{eom} provided the metric functions and scalar field are
\begin{align}\label{solsoliton1}
u(\rho) &= \mu_1 - \frac{\gamma^2}{4\rho^2}-\frac{\Lambda \rho^2}{2}, & v(\rho) &= \frac{\gamma}{2\rho^2}, & \phi(t) &= \lambda_1\,t\,, 
\end{align}
where $\mu_1$, $\gamma$ and $\lambda_1$ are integration constants; the latter satisfying the condition $\lambda_1^2=\Lambda/\kappa$. Notice that, in contrast to the black strings in Sec.~\ref{sec:CSMG}, this solution is supported only by a positive cosmological constant, such that the time-dependent scalar field remains real. The Lorentzian signature of this solution is preserved provided $\mu_1>\sqrt{\Lambda\gamma^2/2}$ and the range for the radial coordinate is restricted to $\rho\in[\rho_-,\rho_+]$, where
\begin{equation}\label{rpmsoliton1}
    \rho_{\pm}=\left[\frac{\mu_1\pm\sqrt{\mu_1^2-\Lambda\gamma^2/2}}{\Lambda}\right]^{1/2}\ .
\end{equation}
The absence of conical singularities demands that the $z$-coordinate has to be identified as $z\sim z+\beta_z^{(\pm)}$ at $\rho_\pm$, whose period is $\beta_z^{(\pm)}=4\pi/u'(\rho_\pm)$. This, in turn, implies that the periodicity of the angular coordinate is $\varphi\sim\varphi+\beta_\varphi^{(\pm)}$, where $\beta_\varphi^{(\pm)}=-\beta_z^{(\pm)}v(\rho_\pm)$.

It is clear that this configuration is a product space $\mathbb{R}\times M^3$, where $M^3$ is compact (as displayed above) and has positive definite geometry. Moreover, a straightforward calculation shows that $M^3$ has positive curvature. Then, a theorem by Richard S. Hamilton implies that $M^3$ is diffeomorphic to $S^3$, considering that we want spacetime to be simply-connected, see~\cite{Ziller} and references therein. Indeed, one can check that it is a non-Einstein conformally flat spacetime supported by a time-dependent scalar field. This can be seen by computing its traceless Ricci and Weyl tensor, respectively, and noticing that the former is nonvanishing while the latter is zero.

In summary, by using the Cardy formula we have established $\mbox{AdS}_3\times\mathbb{R}$ as the ground state of the black string. We have found that both $\mbox{AdS}_3\times\mathbb{R}$ and $\mathbb{R}\times S^3$ are non-trivial solutions of CSMG. These spaces are conformally flat non-Einstein product spaces sourced by nontrivial scalar fields. These spaces are analogs of the Pleba\'nski-Hacyan spacetimes, which have geometries $\mbox{AdS}_2\times\mathbb{R}^2$ and $\mathbb{R}^2\times S^2$~\cite{Plebanski:1979}.
However, $\mathbb{R}\times S^3$ is incompatible with the symmetries of the black string and they do not share the same boundary conditions. This implies that they cannot be compared as configurations that belong to the same ensemble. We postpone a deeper study of its properties for future investigations. Thus, from hereon, we only consider the ground state in a thermodynamic context.

\section{Thermodynamic Stability and Phase Transitions \label{sec:local}}

It is well known that the second law of thermodynamics is equivalent to the weakly-convex property of the internal energy function~\cite{Callen}. In our framework, what is relevant is to analyze the concavity of the free energy $\mathcal{G}(T,\Omega)$. Phase transitions occur when this function is not analytic. A function with multiple branches corresponds to a system with various phases. The branch of the function which has less free energy is statistically preferred and it represents a stable state of the system under fluctuations.

From previous sections, we know that the free energy of the black string is
\begin{equation}\label{Gibbs}
\mathcal{G}(T,\Omega)=-\frac{2\pi^{2}l^{2}\sigma T^{2}}{\kappa(1-\Omega^{2}l^{2})}\ .
\end{equation}
Notice that it satisfies the concavity conditions
\begin{equation}\label{lsconds}
\frac{\partial^{2} \mathcal{G}}{\partial T^{2}} \leq 0, \quad \frac{\partial^{2} \mathcal{G}}{\partial \Omega^{2}} \leq 0, \quad \frac{\partial^{2} \mathcal{G}}{\partial T^{2}} \frac{\partial^{2} \mathcal{G}}{\partial \Omega^{2}}-\left(\frac{\partial^{2} \mathcal{G}}{\partial T \partial \Omega}\right)^{2} \geq 0 ,
\end{equation}
since the existence of event horizons is determined by $|j|< m l$ which requires $\Omega^2\ell^2<1$.


Now, the local stability of the black string can be analyzed by studying its response under small perturbations of its thermodynamic variables around equilibrium. Conditions~\eqref{lsconds} can be interpreted as manifestations of the response functions. For instance, the specific heat at constant angular velocity, $C_{\Omega}=T{(\frac{\partial S}{\partial T})}_{\Omega}$, measures the heat absorption from a temperature stimulus, while the isothermal compressibility for the angular momentum, $\kappa_T=\frac{1}{J}{(\frac{\partial J}{\partial \Omega})}_{T}$, measures the response of the angular momentum respect to an angular velocity stimulus. Specifically, these quantities are
\begin{align}
    C_{\Omega}=\frac{4\pi^2 l^2 \sigma T}{\kappa(1-\Omega^2 l^2)}\;\;\;\;\; \mbox{and}\;\;\;\;\;\kappa_T=\frac{1+3l^2\Omega^2}{\Omega (1-\Omega^2 l^2)}\ .
\end{align}
The stability conditions for these quantities are $C_{\Omega}\geq 0$ as well as $J \kappa_T\geq 0$. The former is clearly satisfied while the latter is
\begin{equation}
J \kappa_T=\frac{4\pi^2 l^2 (1+3l^2\Omega^2)\sigma T^2}{\kappa(1-\Omega^2 l^2)^3}\geq 0\ .
\end{equation}
In consequence, the analytic rotating black string is always able to reach a thermal equilibrium with a heat bath.\footnote{The extremal case when $|j|=ml$ has $\mathcal{G}=0$ and in consequence it is locally stable}

We have seen  that the action principle of dynamical Chern-Simons modified gravity admits two solutions with the same boundary conditions for a fixed temperature. These are the rotating black string presented in Sec.~\ref{sec:CSMG} and the $\mbox{AdS}_3\times\mathbb{R}$ ground state described by the line element~\eqref{soliton2} with \eqref{solsoliton2}. Both solutions share the same asymptotic geometry and behavior, representing two phases of the same system. Therefore, global stability is analyzed by comparing their respective free energies. For simplicity, we consider only static configurations and emphasize that we require both configurations to be at the same temperature.

As it was pointed out in Sec.~\ref{sec:solitons}, the ground state lacks an event horizon and its temperature can be chosen arbitrarily to match the black string temperature. Thus, from Eqs.~\eqref{M2} and~\eqref{Gibbs} we have
\begin{align}
  \mathcal{G}=-\frac{2\pi^2}{\kappa}\ell^2\sigma T^2   \;\;\;\;\; \mbox{and}\;\;\;\;\; \mathcal{G}_0=M_0=-\frac{\sigma}{2\kappa}\ .
\end{align}
Since $\ell^{-2}=\kappa\lambda^2/2$, then the difference between both Gibbs free energies is
\begin{equation}\label{deltag}
    \Delta \mathcal{G}=\mathcal{G}-\mathcal{G}_0=\left(\frac{2\pi}{\kappa\lambda}\right)^2\left(\frac{\kappa\lambda^2}{8\pi^2}-T^2\right)\sigma\ .
\end{equation}
A phase transition between the static black string and the ground state in Eq.~\eqref{soliton2} occurs at the critical temperature $T_c$, when the condition $\Delta \mathcal{G}=0$ holds, namely,
\begin{equation}\label{Tc}
    T_c= \frac{\lambda}{2\pi}\sqrt{\frac{\kappa}{2}} \,.
\end{equation}
Therefore, a phase transition is supported by the existence of the scalar field, as it can be seen from the linear dependence of the critical temperature on the axionic charge $\lambda$. Moreover, this is a first order phase transition because the slope of $\Delta \mathcal{G}$ does not vanish at $T_c$. Taking into account Eq.~\eqref{deltag}, we see that the ground state is preferred for temperatures below the critial temperature, i.e. $T<T_c$, and the black string is thermodynamically favored when $T>T_c$.


\section{Conclusions\label{sec:conclusions}}

In this work, we study black string spacetimes possessing axionic charge in dynamical CSMG. We calculate their conserved charges and use them to construct a Cardy formula to establish the ground state. We later move on to the thermodynamics of this black string/ground state system, finding phase transitions between the pair. 

The analysis is carry out by considering a scalar field with linear dependence on the coordinate that spans the string. Although this assumption might break the isometry group, the shift symmetry in field space renders the energy-momentum tensor of scalar fields compatible with the symmetries of the metric. Indeed, this class of scalar fields has been widely used in the context of holography, providing simple models with momentum relaxation (see~\cite{Andrade:2013gsa} and references therein). 

First, we review the analytic rotating BTZ black string solution in CSMG found in Ref.~\cite{Cisterna:2018jsx}. It is worth recalling that exact rotating configurations in this theory are scarce since, in general, nontrivial contribution are introduced by the Chern-Simons coupling in the field equations~\cite{Grumiller:2007rv}. Nevertheless, we demonstrate that the presence of the scalar field with axionic profile is crucial for obtaining the locally $\mbox{AdS}_3\times\mathbb{R}$ rotating black string, due to the addition of nontrivial contributions to the off-diagonal components of the field equations. 

Afterwards, we compute conserved charges by means of the renormalized boundary stress-energy tensor, obtaining the mass and angular momentum of the black string. The entropy is obtained through the Wald formalism, by evaluating the Noether charge at the horizon. The area law is recovered despite the fact that we are in an effective gravity theory with higher-curvature corrections. Using the Hamiltonian approach, we obtain the free energy by computing the renormalized Euclidean on-shell action, augmented by boundary terms that guarantee finiteness on asymptotically locally $\mbox{AdS}_3\times\mathbb{R}$ spacetimes while defining a well-posed variational principle. We recover the previous conserved charges as thermodynamic quantities and find that the first law of thermodynamics is satisfied. The axionic charge is fixed in terms of the cosmological constant and it can be only treated as a thermodynamic variable in the extended phase space (see~\cite{Kubiznak:2014zwa}); approach that we do not follow here. Additionally, exploiting the rescaling symmetry along the radial direction, we derive a Smarr formula from the Noether theorem following the approach of Ref.~\cite{BT}. We find that the latter is equivalent to that of the rotating BTZ black hole in three-dimensional gravity.

Then, we obtain the ground state and a conformally flat non-Einstein product space with axionic charge by performing the analytic continuation of the black string along the flow generated by their Killing vector fields. All the solutions obtained through this method are regular and horizonless. Therefore, their lack of entropy implies that their energy can be obtained from the renormalized Euclidean on-shell action with arbitrary value of the Euclidean time. We compute the energy of the ground state explicitly and find that it is negative, representing the Casimir energy in the dual field theory when the fermions are antiperiodic on the compact coordinate~\cite{Coussaert:1993jp}. 

We study thermodynamic stability and phase transitions in Sec.~\ref{sec:local}. The Gibbs free energy of the black string is a concave function of the temperature and angular velocity implying that this configuration is thermodynamically locally stable under small perturbations in temperature and angular momentum. We verify that thermal equilibrium with a heat bath can always be attained by analyzing the response functions such as the heat capacity and the isothermal compressibility of the angular momentum. The global stability is a noteworthy aspect, considering the existence of a ground state in presence of an axionic scalar field. 
In particular, we find that a first-order phase transition takes place at certain critical temperature $T_c$ [see Eq.~\eqref{Tc}]. For $T<T_c$ the gravitational ground state is thermodynamically favored, while for $T>T_c$ the black string becomes the most probable configuration. 

Interesting phenomena can be explored in future works. For instance, a stabilization mechanism of the Gregory--Laflamme instability due to axions in CSMG is worth pursuing, similar to the one proposed in~\cite{Cisterna:2019scr}. This might provide evidence on the dynamical stability of the black string, as it has been recently analyzed in GR~\cite{Henriquez-Baez:2021gdn,Dhumuntarao:2021gdb}. Additionally, the extended phase space could certainly enrich the thermodynamics of phase transitions. This is because the relation between the axionic charge and the cosmological constant may offer a scenario where the former, as a free parameter, is interpreted as a pressure. Finally, a natural extension of these solutions is to include Maxwell fields. To this end, one could implement the approach of Refs.~\cite{Liu:2019rib,Cisterna:2020rkc,Cisterna:2021ckn}, rendering this class of solutions compatible with the presence of electric/magnetic charge.

\begin{acknowledgments}
The authors thank I.~J.~Araya, A.~Argando\~na, A.~Cisterna, O.~J.~Franca, N.~Merino, R.~Olea, M.~Pino,  R.~Troncoso, L.~F.~Urrutia, and O.~Valdivia for valuable comments, suggestions, and insightful remarks. We also acknowledge the collaboration of S.~del Pino in early stages of this project. The work of C. C. is supported by Agencia Nacional de Investigación y Desarrollo (ANID) through FONDECYT No~11200025 and~1210500. C. E. is supported by ANID through PAI Grant No. 77190046 and Universidad Central de Chile through Proyecto Interno CIP2020039. D. F.-A. is supported by a CONACYT postdoctoral fellowship and also acknowledges support from CONACYT Grant No A1-S-11548. K.L. is financially supported by Beca Doctorado Nacional No 21182110, ANID.
\end{acknowledgments}

\bibliography{References}

\end{document}